\documentclass[reqno,a4paper,11pt]{article}
\pdfoutput=1
\usepackage{color}

\usepackage{graphicx}
\usepackage[textwidth = 430 pt, textheight = 630 pt]{geometry}

\definecolor{MyDarkBlue}{rgb}{0.15,0.25,0.45}
\usepackage{epsfig,rotating}
\usepackage{amsmath,amssymb}
\usepackage{amsfonts}
\usepackage{mathrsfs}
\usepackage{bbm}

\usepackage{latexsym}
\usepackage{amsthm}

\usepackage[utf8x]{inputenc}

\usepackage{hyperref}
\hypersetup{
colorlinks=true,
citecolor=MyDarkBlue,
linkcolor=MyDarkBlue,
urlcolor=MyDarkBlue,
pdfauthor={Sam Palmer and Christian S\"amann},
pdftitle={Self-dual String and Higher Instanton Solutions},
pdfsubject={hep-th math-ph}
breaklinks=true
}

\usepackage{tikz}
\usepackage{mathtools}

\theoremstyle{remark}

\theoremstyle{definition}

\theoremstyle{remark}

\let\fn\footnote
\renewcommand{\footnote}[1]{\linespread{1.1}\fn{#1}\linespread{1.29}}

\makeatletter\renewcommand{\section}{\@startsection
{section}{1}{\z@}{-3.5ex plus -1ex minus
    -.2ex}{2.3ex plus .2ex}{\bf }}
\makeatletter\renewcommand{\subsection}{\@startsection{subsection}{2}{\z@}{-3.25ex
plus -1ex minus
   -.2ex}{1.5ex plus .2ex}{\it }}
\makeatletter\renewcommand{\subsubsection}{\@startsection{subsubsection}{3}{-2.45ex}{-3.25ex
plus -1ex minus -.2ex}{1.5ex plus .2ex}{\it }}
\renewcommand{\thesection}{\arabic{section}}
\renewcommand{\thesubsection}{\arabic{section}.\arabic{subsection}}
\renewcommand{\@seccntformat}[1]{\@nameuse{the#1}.~~}

\renewcommand{\theequation}{\thesection.\arabic{equation}}
\makeatletter \@addtoreset{equation}{section}

\usepackage[toc,page]{appendix}

\renewcommand{\appendices}{
\section*{Appendix}\label{appendices}\setcounter{subsection}{0}
\addcontentsline{toc}{section}{Appendix}
\setcounter{equation}{0}
\makeatletter
\renewcommand{\theequation}{\Alph{subsection}.\arabic{equation}}
\renewcommand{\thesubsection}{\Alph{subsection}}
\@addtoreset{equation}{subsection}
\makeatother
}

\def\slasha#1{\setbox0=\hbox{$#1$}#1\hskip-\wd0\hbox to\wd0{\hss\sl/\/\hss}}

\def\periodb#1{\setbox0=\hbox{$#1$}#1\hskip-\wd0\hbox to\wd0{-}}

\newcommand{\unit}{\mathbbm{1}}   			% identity map/matrix
\newcommand{\CA}{\mathcal{A}}    			% cal-letters

\newcommand{\xb}{\bar{x}}
\newcommand{\xh}{\hat{x}}

\newcommand{\CC}{\mathcal{C}}

\newcommand{\CCI}{\mathscr{I}}

\newcommand{\CF}{\mathcal{F}}

\newcommand{\CH}{\mathcal{H}}

\newcommand{\CN}{\mathcal{N}}
\newcommand{\CO}{\mathcal{O}}

\newcommand{\frder}{\mathfrak{der}}				% frak-letters
\newcommand{\frg}{\mathfrak{g}}				% frak-letters
\newcommand{\frh}{\mathfrak{h}}				% frak-letters

\newcommand{\frl}{\mathfrak{l}}

\newcommand{\FR}{\mathbbm{R}}     			% field of real numbers
\newcommand{\FC}{\mathbbm{C}}     			% field of complex numbers
\newcommand{\FH}{\mathbbm{H}}     			% field of quaternions
\newcommand{\NN}{\mathbbm{N}}     			% set of natural numbers
\newcommand{\RZ}{\mathbbm{Z}}     			% ring of integers
\newcommand{\CPP}{{\mathbbm{C}P}}    			% complex projective plane
\newcommand{\dd}{\mathrm{d}}     			% total differential
\newcommand{\dpar}{\partial}     			% partial differential
\newcommand{\di}{\mathrm{i}}     			% imaginary unit
\newcommand{\eps}{{\varepsilon}}			% antisymmetric tensors
\renewcommand{\Im}{\mathrm{Im}}     			% barred letters
\newcommand{\zb}{{\bar{z}}}

\newcommand{\sigmab}{{\bar{\sigma}}}

\newcommand{\eand}{{\qquad\mbox{and}\qquad}}     		% and etc. in equations
\newcommand{\ewith}{{\qquad\mbox{with}\qquad}}

\newcommand{\tr}{\,\mathrm{tr}\,}     			% trace
\newcommand{\au}{\mathfrak{u}}
\newcommand{\asu}{\mathfrak{su}}

\newcommand{\aspin}{\mathfrak{spin}}

\newcommand{\sU}{\mathsf{U}}     			% groups

\newcommand{\sG}{\mathsf{G}}
\newcommand{\sSU}{\mathsf{SU}}

\newcommand{\sO}{\mathsf{O}}

\newcommand{\sSO}{\mathsf{SO}}

\newcommand{\acton}{\vartriangleright}     			% span
\def\tyng(#1){\hbox{\tiny$\yng(#1)$}}			% small Young diagram
\def\tyoung(#1){\hbox{\tiny$\young(#1)$}}			% small Young diagram
\newcommand{\beq}{\begin{eqnarray}}
\newcommand{\eeq}{\end{eqnarray}}

\newcommand{\sft}{{\sf t}}

\newcommand{\sfd}{{\sf d}}

\newcommand{\sff}{{\sf f}}

\newenvironment{conditions}{
\vspace{-2mm}\begin{itemize}
\setlength{\itemsep}{-1mm}
}{\vspace{-2mm}\end{itemize}}

\begin{document}

\begin{titlepage}
\begin{flushright}
 EMPG--13--24
\end{flushright}
\vskip 2.0cm
\begin{center}
{\LARGE \bf Self-dual String and Higher Instanton Solutions}
\vskip 1.5cm
{\Large Sam Palmer and Christian S\"amann}
\setcounter{footnote}{0}
\renewcommand{\thefootnote}{\arabic{thefootnote}}
\vskip 1cm
{\em Maxwell Institute for Mathematical Sciences\\
Department of Mathematics, Heriot-Watt University\\
Colin Maclaurin Building, Riccarton, Edinburgh EH14 4AS, U.K.}\\[0.5cm]
{Email: {\ttfamily sap2@hw.ac.uk~,~c.saemann@hw.ac.uk}}
\end{center}
\vskip 1.0cm
\begin{center}
{\bf Abstract}
\end{center}
\begin{quote}
We present and discuss explicit solutions to the non-abelian self-dual string equation as well as to the non-abelian self-duality equation in six dimensions. These solutions are generalizations of the 't Hooft-Polyakov monopole and the BPST instanton to higher gauge theory. We expect that these solutions are relevant to the effective description of M2- and M5-branes.
\end{quote}
\end{titlepage}

\section{Introduction}

As well as being an interesting field in its own right, studying self-dual strings is a promising avenue of approach to understanding systems of multiple M5-branes. Self-dual strings are configurations of M2-branes ending on M5-branes and only the abelian case involving a single M5-brane is well understood \cite{Howe:1997ue}. A complete non-abelian formulation could yield supersymmetry transformations for the $\CN$=(2,0) theory describing multiple M5-branes, which in turn could fix the underlying equations of motion.

Recent progress in M2-brane models suggests that Lie algebras need to be generalized to capture the gauge structure of effective field theories in M-theory. For example, the description of two M2-branes can be formulated in the BLG model \cite{Bagger:2007jr,Gustavsson:2007vu} based on the 3-Lie algebra $A_4$, while underlying the ABJM model \cite{Aharony:2008ug,Bagger:2008se} are hermitian 3-Lie algebras. Such 3-algebras also seem to be relevant to the study of M5-branes. In particular, they can be regarded as so-called differential crossed modules, which form the generalized gauge algebra of non-abelian gerbes \cite{Palmer:2012ya,Palmer:2013ena}.

The string theory interpretation of monopoles in terms of D1-branes ending on D3-branes explains several key features of the field theory description of monopoles. For example, the numbers of D3- and D1-branes correspond to the rank of the gauge group and the topological charge on the field theory side. Furthermore, the uniqueness of the 't Hooft-Polyakov monopole, up to translations, is captured by the string theory interpretation of a single D1-brane stretched between two D3-branes. We expect similar deep insights from comparing explicit self-dual string solutions to their M-brane interpretation.

In this paper, we present explicit solutions to the self-dual string equation arising in the context of higher gauge theory. In particular, we consider a spherically symmetric ansatz that is a rather straightforward generalizations of the 't Hooft-Polyakov monopole. As we expect a close link to M2-brane models, we base our ansatz on the differential crossed module corresponding to $A_4$, the 3-Lie algebra appearing prominently in the BLG model. It turns out that this ansatz can be solved, and the scalar field of the self-dual string configuration can be classified by integer winding numbers, just as the scalar field of the $\sSU(2)$ monopole.

Further motivation for the study of elementary self-dual string solutions stems from our goal to establish an ADHMN-like construction of self-dual strings. Such a construction had been developed using loop spaces in \cite{Saemann:2010cp,Palmer:2011vx}, but the corresponding picture in higher gauge theory remains unknown. The related twistor constructions, however, were given in \cite{Saemann:2012uq,Saemann:2013pca}, making it reasonable to expect the existence of such a construction.

Given a potential ADHMN-like construction of self-dual strings, it is only natural to ask for an analogue of the ADHM construction, which would yield solutions to the self-duality equations in six dimensions. For lack of a better name, we will call such solutions higher instantons. Again, a twistor description of higher instantons was given in \cite{Saemann:2012uq,Saemann:2013pca}. To develop an ADHM-like construction, a good understanding of the elementary solutions to the higher instanton equation is crucial.

Using an ansatz closely related to the BPST instanton, we manage to find explicit higher instanton solutions which can be continued to solutions on a large region in the conformal compactification of six-dimensional Minkowski space. In fact, our solutions are invariant under an action of $\sSO(1,5)$ and share many of the properties of the BPST instanton.

Having presented our solutions, we discuss in detail the gauge transformations and fake curvature conditions of higher gauge theory. This is a subtle point, and our analysis suggests to switch to differential 2-crossed modules for a complete picture.

This paper is structured as follows. In section 2, we review the basic monopole and instanton solutions as well as the properties we wish to recover in the higher gauge theoretic setting. In section 3, we discuss our self-dual string solutions and the higher instanton solutions are given in section 4. We conclude in section 5. Two appendices recall the definitions of 3-algebras and differential crossed and 2-crossed modules for the reader's convenience.

\section{Monopoles and instantons}\label{sec:MI}

In the following, we give a concise review of monopoles and instantons from both the field and string theory perspectives. We also quote the simplest explicit solutions, which will serve as inspiration for our ans\"atze for self-dual string and higher instanton solutions.

\subsection{Monopoles}

Monopoles on $\FR^3$ are defined as solutions to the {\em Bogomolny equation}\footnote{For simplicity we set the electric charge to $e=1$.}
\begin{equation}\label{eq:Bogomolny}
F:=\dd A+\tfrac{1}{2}[A,A]=\star(\dd \Phi + [A,\Phi]) ~,
\end{equation}
where the connection one-form $A$ and the function $\Phi$ take values in a Lie algebra $\frg$ and the asymptotic behavior of the field $\Phi$ is
\begin{equation}\label{eq:fall-off}
|\Phi(x)|=v-\frac{q}{|x|}+\CO\left(\frac{1}{|x|^{2}}\right) ~~~\mbox{as}~~~|x|\rightarrow\infty
\end{equation}
for some $v\in\FR$ and $q\in\RZ$. Here, $|x|=\sqrt{x^ix^i}$ on $\FR^3$ and $|\Phi(x)|$ is defined using the Killing form on $\frg$. 

If the gauge group is $\sSU(2)$, we can use the asymptotic behavior $|\Phi|\sim v$ to impose the following asymptotic gauge condition on $\Phi$:
\begin{equation}\label{eq:su(2)Higgs}
\Phi\sim \gamma^{-1}\left(\begin{array}{cc} v & 0\\ 0& -v \end{array}\right) \gamma~.
\end{equation}
The elements $\gamma\in\sSU(2)$ which leave this expression invariant form the stabilizing group $\sU(1)$. Solutions are therefore classified by an integer topological charge
\begin{equation}
\pi_2(\sSU(2)/\sU(1))\cong\RZ~.
\end{equation}
This integer is given by $q$ in \eqref{eq:fall-off} and it is called the {\em charge of the monopole}. This charge can be computed alternatively as
\begin{equation}\label{eq:top_charge_monopole}
 2\pi q=\tfrac{1}{2}\int_{S^2_\infty}\frac{\tr(F^\dagger \Phi)}{||\Phi||}\ewith ||\Phi||:=\sqrt{\tfrac{1}{2}\tr(\Phi^\dagger \Phi)}~,
\end{equation}
where the integral is taken over the sphere at infinity, $S^2_\infty$.

In string theory, monopoles of charge $q$ with gauge algebra $\au(N)$ correspond to stacks of $q$ D1-branes ending on stacks of $N$ D3-branes in type IIB superstring theory as follows  \cite{Diaconescu:1996rk}:
\begin{equation}\label{diag:D1D3}
\begin{tabular}{rcccccc}
& 0 & 1 & 2 & 3 & 4  & \ldots\\
D1 & $\times$ & & & & $\times$ \\
D3 & $\times$ & $\times$ & $\times$ & $\times$ &
\end{tabular}
\end{equation}
This is a BPS configuration and the corresponding time-independent BPS equation in the low-energy effective description of the D3-branes is the Bogomolny equation \eqref{eq:Bogomolny}. The field $\Phi$ describes fluctuations of the D3-branes in the $x^4$ direction, i.e.\ parallel to the D1-branes. If the field $\Phi$ is singular, the poles represent points from which the D1-branes stretch to infinity. Non-singular monopoles, on the other hand, correspond to D1-branes stretched between D3-branes, and therefore must be non-abelian. In particular, a monopole configuration with a scalar field as in \eqref{eq:su(2)Higgs} corresponds to D1-branes suspended between parallel D3-branes separated by a distance $2v$.

\subsection{Basic monopole solutions}

The simplest monopole is the {\em Dirac monopole} with $N=1$, $q=1$. It is unique up to translations $x\rightarrow x+x_0$ and the field configuration reads
\begin{equation}\label{eq:Dirac}
\begin{aligned}
\Phi&=\di v+\frac{\di}{|x|}~,~~~A_\pm=\frac{\di}{2r}\left(\frac{z_\pm~\dd \zb_\pm}{1+|z_\pm|^2}-\frac{\zb_\pm~\dd z_\pm}{1+|z_\pm|^2}\right)~,\\
F&=\di\eps_{ijk} \frac{x^k}{|x|^3}~\dd x^i\wedge\dd x^j=\frac{\di}{|r|^2}~ \mbox{vol}_{S^2}=\frac{\di}{r^2}~ \frac{\dd z_\pm\wedge\dd \zb_\pm}{(1+|z_\pm|^2)^2}~,
\end{aligned}
\end{equation}
where $r$ and $z_\pm$ are the radial and the usual stereographic complex coordinates appearing in the foliation of $\FR^3$ by two-spheres. Besides the singularity in $\Phi$ at the origin, $A_\pm$ are singular along the negative and positive $x^3$-axis, respectively. This singularity is known as the {\em Dirac string}. The configuration \eqref{eq:Dirac} can be scaled by $q\in\NN$ to give $q$ coincident Dirac monopoles. Recall that the monopole connection defines a connection on a principal $\sU(1)$-bundle over a sphere encircling the Dirac monopole. The monopole charge is then the first Chern number of this connection: $q=-\tfrac{\di}{2\pi}\int_{S^2} F$.
 
The Dirac monopole can be embedded into $\sSU(2)$ to give a non-abelian configuration known as the {\em Wu-Yang monopole} \cite{Wu:1976ge}:
\begin{equation}
\begin{aligned}
\Phi&=\left(v-\frac{1}{|x|}\right)\frac{e_ix^i}{|x|}~,~~~A=\eps_{ijk}\frac{e_ix^j}{|x|^2}\dd x^k~,\\
F&=\eps_{ijk} ~x^k~\frac{ e_l x^l}{|x|^4}~\dd x^i\wedge\dd x^j~.
\end{aligned}
\end{equation}
Here, the $e_i$ are generators of $\asu(2)$, defined as $e_i:=\tfrac{\di}{2}\sigma_i$ in terms of Pauli matrices. Again, there is a  singularity at the origin, but the Dirac string is removed. This solution can be further embedded into $\sSU(N)$ and extended to arbitrary charge $q$.

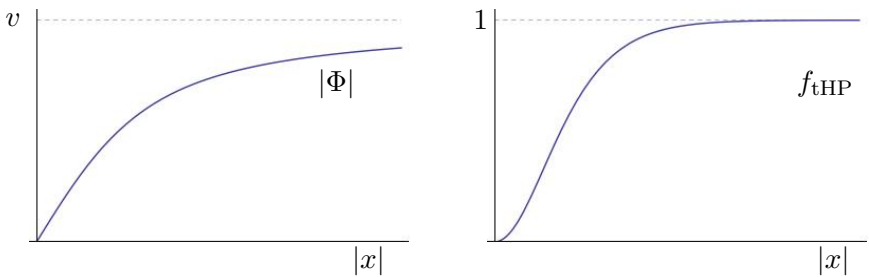
\begin{figure}[h]
\center
\begin{picture}(380,100)
~~~~~~\includegraphics[width=50mm]
{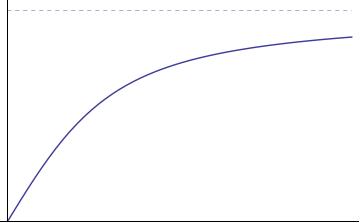}~~~~~~~~\includegraphics[width=50mm]{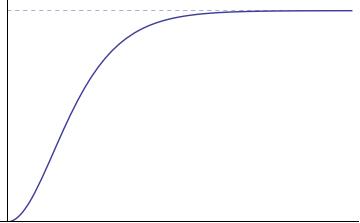}
\put(-187.0,-7.0){\makebox(0,0)[c]{$|x|$}}
\put(-320.0,84.0){\makebox(0,0)[c]{$v$}}
\put(-145.0,84.0){\makebox(0,0)[c]{$1$}}
\put(-13.0,-7.0){\makebox(0,0)[c]{$|x|$}}
\put(-199.0,60.0){\makebox(0,0)[c]{$|\Phi|$}}
\put(-16.0,60.0){\makebox(0,0)[c]{$f_{\rm tHP}$}}
\end{picture}
\caption{The radial dependence of the scalar field $\Phi$ and the function $f_{\rm tHP}(\xi)=(1-\frac{\xi}{\sinh(\xi)})$ appearing in the gauge potential $A$ of the 't Hooft-Polyakov monopole \eqref{eq:thooft}.}
\label{fig:monopoles}
\end{figure}

The 't Hooft-Polyakov monopole \cite{Prasad:1975kr} is the unique $N=2,~q=1$ monopole on $\FR^3$. It reads explicitly as
\begin{equation}\label{eq:thooft}
\begin{aligned}
\Phi&=\frac{e_i x^i }{|x|^2}\big(\xi\ {\rm coth}(\xi)-1\big)~,\\
A&=\eps_{ijk}\frac{e_i x^j}{|x|^2}\left(1-\frac{\xi}{\sinh(\xi)}\right)~\dd x^k~,
\end{aligned}
\end{equation}
where $\xi:=v|x|$ is dimensionless and $e^i$ are again the generators of $\asu(2)$. The topological charge of this solution is $q=1$, as one readily computes using \eqref{eq:top_charge_monopole}. Note that in this solution, all singularities are removed, cf.\ figure \ref{fig:monopoles}. However, this solution cannot be extended to higher $q$ and it is the only spherically symmetric non-singular monopole \cite{Rossi:1982fq} with gauge group $\sSU(2)$. 

\subsection{Instantons}

Instantons on $\FR^4$ are defined as solutions to the self-duality equation
\begin{equation}\label{eq:instanton}
F=\star F~,
\end{equation}
where the non-abelian curvature $F:=\dd A+\tfrac{1}{2}[A,A]$ takes values in the Lie algebra $\frg$ of some gauge Lie group $\sG$ and vanishes sufficiently rapidly as $|x|\rightarrow\infty$. That is, the curvature becomes pure gauge
\begin{equation}
A\sim \gamma^{-1}\dd \gamma
\end{equation}
as $|x|\rightarrow\infty$ for some $\gamma\in \CC^\infty(\FR^4\backslash \{0\},\sG)$. The function $\gamma$ then defines a map $S^3_\infty \rightarrow \sSU(2)$ with an integer winding number
\begin{equation}
\pi_3(\sSU(2))\cong\RZ~.
\end{equation}
This integer is the {\em instanton number}, which is given by the second Chern number
\begin{equation}\label{eq:instanton_number}
 q=\frac{1}{8\pi^2}\int_{\FR^4} \tr(F^\dagger\wedge F)~.
\end{equation}

Just like monopoles, instanton solutions find a nice interpretation in terms of D-brane configurations. A $q$-instanton with gauge group $\sU(N)$ corresponds to a BPS-configuration of $q$ D0-branes bound to $N$ D4-branes:
\begin{equation}\label{diag:D0D4}
\begin{tabular}{rcccccccc}
& 0 & 1 & 2 & 3 & 4 &  \ldots\\
D0 & $\times$ & & & & \\
D4 & $\times$ & $\times$ & $\times$ & $\times$ &  $\times$
\end{tabular}
\end{equation}
Note that the Bogomolny monopole equation arises from the instanton equation via dimensional reduction. Analogously, this D-brane configuration yields the monopole D-brane configuration \eqref{diag:D1D3} via a T-duality along $x^4$.

\subsection{Basic instanton solutions}

There are no abelian instantons. This is due to the fact that the fall-off conditions on the gauge potential correspond to a continuation of the instanton configuration from $\FR^4$ to $S^4$. The gauge potential then is the local description of the connection on a principal fiber bundle over $S^4$. Such a bundle is characterized by transition functions on the overlap of the two standard patches on $S^4$, which is contractible to an $S^3$. The transition functions are therefore given by elements of $\CC^\infty(S^3,\sU(1))$ or $\pi_3(S^1)$, which are all trivial. Alternatively, one can readily show that the instanton number \eqref{eq:instanton_number} for an abelian instanton necessarily vanishes.

Let us therefore turn to gauge group $\sSU(2)$. Just as the two-sphere $S^2\cong \CPP^1$ is conveniently described by the usual complex stereographic coordinates, the four-sphere $S^4\cong \FH P^1$ is described by analogous quaternionic stereographic coordinates. In the following, we use the notation
\begin{equation}
\begin{aligned}
x=x^i\sigma_i-\di x^4\unit_2\eand\xb=x^\mu \sigma^\dagger_\mu=x^i \sigma_i+\di x^4\unit_{2}~,
\end{aligned}
\end{equation}
where besides their interpretation as quaternion generators, $\sigma_\mu=(\sigma_i,-\di \unit_{2})$ are the van-der-Waerden symbols appearing in the Clifford algebra of $\FR^4$, which is generated by
\begin{equation}
\begin{aligned}
\gamma_\mu=\left(\begin{array}{cc} 0 &  \sigma^\dagger_\mu\\ \sigma_\mu & 0 \end{array}\right)~.
\end{aligned}
\end{equation}

The BPST instanton \cite{Belavin:1975:85,Atiyah:1979iu}, in regular Landau gauge, reads as
\begin{subequations}\label{eq:BPST_solution}
\begin{equation}
\begin{aligned}
A=\Im\left(\frac{x~\dd \xb}{\rho^2+|x|^2}\right)=\frac{1}{2}\left(\frac{x~\dd \xb}{\rho^2+|x|^2}-\frac{\xb~\dd x}{\rho^2+|x|^2}\right)~,
\end{aligned}
\end{equation}
where $\rho$ is a parameter corresponding to the distance of the D0-brane from the D4-brane\footnote{Taking the D0-brane infinitely far away gives a singular configuration known as the `small instanton'. Inversely, bringing the D0-brane into the worldvolume of the D4-brane yields vanishing curvature and thus no instanton.} and $\Im(M)$ denotes the antihermitian part of a matrix $M$. This gauge potential has the $\asu(2)$-valued curvature
\begin{equation}
\begin{aligned}
F=\rho^2\frac{\dd x\wedge\dd \xb}{(\rho^2+|x|^2)^2}~.
\end{aligned}
\end{equation}
\end{subequations}
Similarly the basic anti-instanton, with charge $q=-1$, has curvature
\begin{equation}
\begin{aligned}
F=\rho^2\frac{\dd \xb\wedge\dd x}{(\rho^2+|x|^2)^2}
\end{aligned}
\end{equation}
and satisfies $F=-\star F$.

Note that the formulas for the gauge potential and its curvature are related to those of the Dirac monopole \eqref{eq:Dirac} by setting $\rho=1$ and replacing quaternionic coordinates by complex stereographic coordinates.

\section{Self-dual strings}

\subsection{Abelian self-dual strings}

The monopole D-brane configuration \eqref{diag:D1D3} can be lifted to M-theory, where we first apply a T-duality, to give a D2-D4 brane system, before performing the M-theory lift. After relabeling coordinates, the resulting configuration is
\begin{equation}\label{diag:M2M5}
\begin{tabular}{rccccccc}
${\rm M}$ & 0 & 1 & 2 & 3 & \phantom{(}4\phantom{)} & 5 & 6 \\
M2 & $\times$ & & & & & $\times$ & $\times$ \\
M5 & $\times$ & $\times$ & $\times$ & $\times$ & $\times$ & $\times$ 
\end{tabular}
\end{equation}
which is again BPS. Contrary to the case of monopoles, the corresponding BPS equation in the effective description of M5-branes is well-established only for a single M5-brane, i.e.\ for $N=1$. This is the so-called {\em self-dual string equation} \cite{Howe:1997ue}
\begin{equation}\label{eq:SelfDualString}
H:=\dd B=\star\dd\Phi~,
\end{equation}
where $B$ is a two-form potential with curvature $H$ and $\Phi$ is a scalar field. All fields are $\au(1)$-valued and live on $\FR^4$. The field $\Phi$ is required to exhibit the asymptotic behavior
\begin{equation}\label{eq:asymptotics_M}
 |\Phi(x)|=v-\frac{q}{|x|^2}+\CO\left(\frac{1}{|x|^{3}}\right) ~~~\mbox{as}~~~|x|\rightarrow\infty~,
\end{equation}
where again $v\in \FR$ and $q\in\RZ$. 

The solution analogous to the Dirac monopole \eqref{eq:Dirac} is the Howe-Lambert-West (HLW) self-dual string \cite{Howe:1997ue}
\begin{equation}\label{eq:HLW}
\Phi=\frac{\di}{|x|^2}~,~~H=\frac{\di}{|x|^3}~\mbox{vol}_{S^3}=\di\eps_{\mu\nu\kappa\lambda} \frac{x^\lambda}{|x|^4} \dd x^\mu\wedge \dd x^\nu\wedge \dd x^\kappa~,
\end{equation}
which is a straightforward generalization of the Dirac monopole to four dimensions. It can be rescaled to yield a solution of $q$ M2-branes ending on $N=1$ M5-branes. The charge is here the Dixmier-Douady class of an abelian gerbe on an $S^3$ encircling the position of the self-dual string in $\FR^4$, $q=-\frac{\di}{2\pi}\int_{S^3} H$.

Another interesting abelian configuration is the Perry-Schwarz (PS) self-dual string \cite{Perry:1996mk}. This is a solution to a non-linear self-dual string equation, which is non-BPS. Interestingly, the solution is non-singular and has the asymptotic behavior \eqref{eq:asymptotics_M} expected of a non-abelian self-dual string. There is in fact a whole family of solutions to the PS self-dual string equation with an $\sO(1,1)$-symmetry, suggesting an underlying string theory description  \cite{Berman:2003xz}. This family interpolates between the PS self-dual string and the HLW self-dual string, which is also a solution to the non-linear equation.

\subsection{Non-abelian self-dual strings}

There are various proposals for a non-abelian generalization of the self-dual string equation \eqref{eq:SelfDualString}, which should describe configurations involving $N\ge 2$ M5-branes. In this section, we review the equations arising in the context of higher gauge theory and compare them to other recent proposals.

M5-branes interact via M2-branes ending on them. The M2-brane boundaries are called self-dual strings, and an effective description of such systems should involve the parallel transport of these self-dual strings. Parallel transport of extended objects is captured by higher gauge theory, which is the theory of non-abelian gerbes with connective structure. Even though we partially fixed the worldvolume of the self-dual string in the above configuration to fill the $x^5$-direction, we still expect that the relevant description originates from higher gauge theory.

In particular, we start from a pair of Lie algebras $\frh$ and $\frg$ forming a differential crossed module\footnote{See appendix \ref{app:diff} for the relevant definitions.}, which takes over the role of the gauge algebra in higher gauge theory. The local connective structure of a non-abelian gerbe on $\FR^4$ is then given by a potential one-form $A\in \Omega^1(\FR^4,\frg)$ and a potential two-form $B\in\Omega^2(\FR^4,\frh)$. The non-abelian scalar field takes values in $\frh$. The potential one-form gives rise to a connection $\nabla=\dd+A$, which can act on both $B$ and $\Phi$ via the action $\acton:\frg\times \frh\rightarrow \frh$ included in the definition of a crossed module. The non-abelian self-dual string equation then reads as 
\begin{equation}\label{eq:NonAbSelfDualString}
H:=\dd B+A\acton B=\star(\dd\Phi+A\acton \Phi)~,
\end{equation}
which was first suggested in \cite{Saemann:2012uq}, where also a construction mechanism for solutions was developed using a twistor approach. In the canonical description of higher gauge theory, the so-called \emph{fake curvature condition}
\begin{equation}\label{eq:fc}
\begin{aligned}
\CF:=\dd A+\tfrac{1}{2}[A,A]-\sft(B)=0
\end{aligned}
\end{equation}
is imposed. This equation guarantees that the parallel transport is consistent and it eliminates additional degrees of freedom from the potential one-form. In the following, we will not impose the fake curvature condition until we return to a more detailed discussion of this issue in section \ref{sec:fakecurvature}.

The infinitesimal gauge transformations which leave the self-dual string equation \eqref{eq:NonAbSelfDualString} together with the fake curvature condition \eqref{eq:fc} invariant read as
\begin{equation}\label{eq:Space-time-GT}
\delta A=\dd  \alpha+[A,\alpha]-\sft(\Lambda)~,~~~
\delta B=\dd  \Lambda +A\acton \Lambda-\alpha\acton B~,~~~
\delta \Phi=-\alpha\acton \Phi~,
\end{equation}
where $\alpha\in \CC^\infty(\FR^4,\frg)$ and $\Lambda\in \Omega^1(\FR^4,\frh)$ are the gauge parameters. Gauge transformations for which $\Lambda$ or $\sft(\Lambda)$ vanish are known as \emph{thin} or \emph{ample} gauge transformations, respectively, as opposed to the general, {\em fat} gauge transformations.

As a further generalization (and categorification), these fields could take values in a differential 2-crossed module, see appendix \ref{app:diff} for definitions. A twistor construction of self-dual strings involving differential 2-crossed modules was presented in \cite{Saemann:2013pca}. The equations of motion arising from this twistor construction are more complicated but can be gauge fixed to the equations above, with only thin gauge transformations remaining.

Differential crossed modules are equivalent to strict Lie 2-algebras. Generalizing to the semistrict case, we obtain 2-term $L_\infty$-algebras, see $\cite{Baez:2002jn}$. The effect of this for the self-dual string equation would be an additional term 
\begin{equation}\label{eq:semi}
H:=\dd B+\mu_2(A, B)+\tfrac{1}{3!}\mu_3(A,A,A)=\star(\dd\Phi+\mu_2(A, \Phi))~,
\end{equation}
where $\mu_i$ are antisymmetric maps satisfying the homotopy Jacobi identities of the $L_\infty$-algebra.

For the special case of a differential crossed module corresponding to a 3-Lie algebra, the self-dual string equation \eqref{eq:NonAbSelfDualString} arose as the BPS equation in the Lambert-Papageorgakis $\CN=(2,0)$ model \cite{Lambert:2010wm}. This model came with an additional vector field $C^\mu$. A self-dual string solution for this model should also satisfy the equations of motion
\begin{equation}\label{eq:(2,0)eom}
\nabla^\mu\nabla_\mu\Phi=0~,~~D(H_{\mu\nu\kappa},C^{\kappa})=F_{\mu\nu}~,~~\nabla_\mu C^\nu=D(C^\mu,C^\nu)=C^\mu\nabla_\mu\Phi=C^\mu\nabla_\mu H_{\nu\kappa\lambda}=0~,
\end{equation}
where $D$ is a map $\frh\wedge\frh\rightarrow\frg$, cf.\ appendix \ref{app:3algebras}.

Another equation arises from the (1,0) superconformal models of \cite{Samtleben:2011fj} derived from the non-abelian generalization of supersymmetric tensor hierarchies. Rather than in Lie algebras, the fields live in vector spaces endowed with maps similar to Lie brackets satisfying identities which generalize the Jacobi identity, see \cite{Samtleben:2011fj,Palmer:2013pka}. These models contain an additional gauge potential three-form $C$, living in a third vector space. Putting these to zero, the BPS equation in the (1,0) superconformal model reduced to $\FR^4$ then reads as
\begin{equation}\label{eq:(1,0)BPS}
H:=\dd B+ 2\sfd(A,\sft(B))+\sfd(A,\dd A-\tfrac{1}{3}\sff(A,A))=\star(\dd\Phi+2\sfd(A,\sft(\Phi))~,
\end{equation}
where $\sfd$ is a symmetric map $\sfd:\frg\odot\frg\rightarrow\frh$ and $\sff$ is an antisymmetric map $\sff:\frg\wedge\frg\rightarrow\frg$. The equation of motion, which is not implied by \eqref{eq:(1,0)BPS}, is 
\begin{equation}\label{eq:(1,0)eom}
\nabla^2\Phi=\star\sfd(\CF,\star\CF)~,
\end{equation}
where $\nabla^2:=\star\nabla\star\nabla$ with $\nabla:=\dd+2\sfd(A,\cdot)$ and $\CF=\dd A-\tfrac{1}{2}\sff(A,A)+\sft(B)$.

As has been shown in \cite{Palmer:2013pka}, the (1,0) superconformal models have a large overlap with higher gauge theory. In particular, equation \eqref{eq:(1,0)BPS} agrees with \eqref{eq:NonAbSelfDualString} or \eqref{eq:semi} for the right choice of vector spaces and brackets. The solutions presented in this paper therefore also yield solutions to \eqref{eq:(1,0)BPS}. We will comment on this further in section \ref{ssec:comments}. For a special class of gauge structures, the (1,0) equations obtained from tensor hierarchies reduce to the equations proposed independently in \cite{Chu:2011fd}, cf.\ \cite{Palmer:2013pka}. 

In another approach \cite{Chu:2012um}, one direction is singled out (as is common in many descriptions of M5-branes) and an additional relation connecting the curvature and potential two-forms is imposed, which is strongly reminiscent of the fake curvature condition. In the case of self-dual strings, this reads as
\begin{equation}\label{eq:GGconstraint}
F_{ij}=c\int\dd x^4 \dpar_4 B_{ij}~,
\end{equation}
where $i,j=1,\ldots,3$ and $c\in\FR$ is some fixed constant.

All fields live in the same Lie algebra\footnote{or a differential crossed module of the form $\frg\overset{\sft}{\rightarrow}\frg$} and the self-dual string equation reads
\begin{equation}\label{eq:GGeqn}
H:=\dd B+[A, B]=\star(\dd\Phi+[A, \Phi])~.
\end{equation}
This equation is invariant under the gauge transformations
\begin{equation}
\delta A=\dd  \alpha+[A,\alpha]~,~~~\delta B=\Sigma-[\alpha, B]~,~~~\delta \Phi=-[\alpha, \Phi]~,
\end{equation}
where $\Sigma$ is a two-form satisfying $\dd\Sigma+[A,\Sigma]=0$. 

Finally in \cite{Saemann:2010cp,Palmer:2011vx}, a transgression of the self-dual string equation to the loop space of $\FR^4$ was considered. It was shown that a Nahm-like transform can be constructed, which maps solutions to the Basu-Harvey equation to solutions to the transgressed self-dual string equation. Here, all ingredients of the construction reduce to those of the ordinary Nahm construction after imposing the usual M2-brane Higgs mechanism. 

\subsection{Previously constructed solutions}

Before presenting our solutions, we briefly comment on solutions to the equations \eqref{eq:(1,0)BPS}, \eqref{eq:GGeqn} and the loop space self-dual string equation that had been given previously.

First, a loop space self-dual string solution remarkably similar to the 't Hooft-Polyakov monopole was found in \cite{Papageorgakis:2011xg}. This solution has gauge algebra $\asu(2)\times \asu(2)$ and we will see this algebra feature prominently in our new self-dual string solutions below. One issue with the loop space solutions is that the space-time role of the scalar field $\Phi$ remained rather unclear.

In \cite{Akyol:2012cq,Akyol:2013ana}, solutions to the tensor hierarchy BPS equations \eqref{eq:(1,0)BPS} had been constructed. In the solutions corresponding to self-dual strings, however, the $B$-field was always put to zero. The explicit solution given in \cite{Akyol:2012cq} contains a $\au(1)$-valued scalar field $\Phi$ and an $\asu(2)$-valued one-form potential $A$. The solution is $\sSO(4)$-invariant and everywhere regular. Also, similarly to the 't Hooft-Polyakov monopole, the potential one-form $A$ can be gauged away at large radius by turning on a potential two-form $B$ and leaving the abelian Howe-Lambert-West self-dual string \eqref{eq:HLW}.

Solutions to \eqref{eq:GGeqn} similar to Wu-Yang monopoles were constructed in \cite{Chu:2012rk}. These solutions were interpreted as corresponding to $N=2$ M2-branes and were generalized to the case $N>2$ in \cite{Chu:2013hja}, where all fields took values in $\asu(N)$. This class of solutions passes certain consistency checks, in particular the M2-brane spike profiles match supergravity predictions \cite{Chu:2012rk,Chu:2013hja}. These solutions, however, remain singular at the position of the self-dual string.

In \cite{Chu:2013gra}, a construction algorithm was given that turned an $\asu(N)$ monopole solution into a solution to the equations \eqref{eq:GGeqn}. The solution constructed from the 't Hooft-Polyakov monopole is a unit charge non-singular self-dual string, but lacks $\sSO(4)$ invariance. The construction also involved choosing a function with certain asymptotic behavior. In this sense the solution is not unique. This situation is similar to our non-singular and $\sSO(4)$-invariant self-dual string solution presented in the following section.

\subsection{$A_4$ non-singular self-dual strings}\label{ssec:A4solSDS}

We now come to a generalization of the 't Hooft-Polyakov monopole to a self-dual string solution based on differential crossed modules. The first issue here is to find the pair of Lie algebras describing our solution. The scalar field of the 't Hooft-Polyakov solution itself, $\Phi=e_i x^i f(r)$, where $f$ is some radial function, suggests a four-dimensional vector space with basis $e_\mu$, allowing for a scalar field $\Phi\propto e_\mu x^\mu$ for the self-dual string. Recall that from the string theory point of view, the self-dual string equation should be dual to the Basu-Harvey equation \cite{Basu:2004ed}, and one might expect that both equations are related. Evidence for this was given in \cite{Palmer:2012ya}, see also \cite{Palmer:2013ena}. In particular, the Basu-Harvey equation in its simplest form is based on the 3-Lie algebra $A_4$, which turns out to be a crossed module $\frh\xrightarrow{\sft}\frg$ of the form $\FR^4\xrightarrow{\sft=0}\asu(2)\times\asu(2)$. Moreover, since $\dim(\frh)=4$, it is an excellent candidate for the gauge structure of a charge-one self-dual string solution.

The 3-Lie algebra $A_4$ is defined as a four dimensional real vector space endowed with the ternary bracket
\begin{equation}\label{eq:A4}
[e_\mu,e_\nu,e_\rho]=\eps_{\mu\nu\rho\sigma}e_\sigma
\end{equation}
on the basis elements $e_\mu\in A_4$. The Lie algebra of inner derivations is $\frg_{A_4}\cong\asu(2)\times\asu(2)$ and it is the linear span of the derivations $D(a,b)$ with
\begin{equation}
D(a,b)\acton c:=[a,b,c]~,~~a,b,c\in A_4~.
\end{equation}

Having fixed the gauge structure, it remains to make an $\sSO(4)$-invariant ansatz for a solution to the self-dual string equation \eqref{eq:NonAbSelfDualString}. Inspired by the $\sSO(3)$-invariant 't Hooft-Polyakov monopole solution \eqref{eq:thooft}, we set
\begin{equation}\label{eq:SDS}
\begin{aligned}
\Phi&=\frac{ e_\mu x^\mu }{|x|^3}~f(\xi)~,\\
B_{\mu\nu}&=\eps_{\mu\nu\kappa\lambda}\frac{e_\kappa x^\lambda}{|x|^3}~g(\xi)~,\\
A_\mu&= \eps_{\mu\nu\kappa\lambda}  D(e_\nu,e_\kappa) ~\frac{x^\lambda}{|x|^2} ~h(\xi)~,
\end{aligned}
\end{equation}
where $\xi:=v|x|^2$ is a dimensionless parameter, $e_\mu$ are the generators of $\frh=A_4$ and $D(e_\mu,e_\nu)\in \frg=\frg_{A_4}$ are inner derivations. We will now seek solutions with non-singular $|\Phi(x)|$ and asymptotic behavior \eqref{eq:asymptotics_M}.

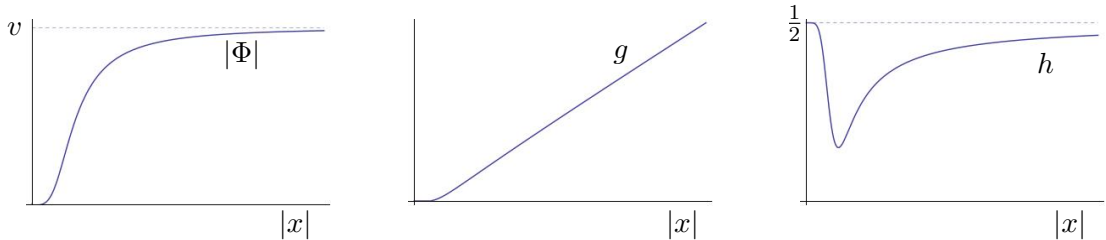
\begin{figure}[h!]
\center
\begin{picture}(400,100)
\includegraphics[width=40mm]
{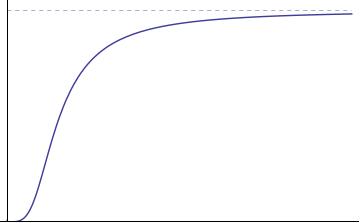}~~~~~~~~\includegraphics[width=40mm]{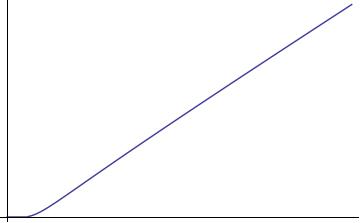}
~~~~~~~~\includegraphics[width=40mm]{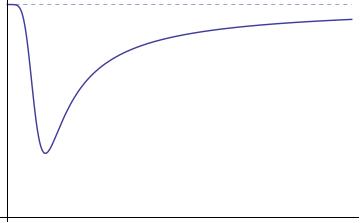}
\put(-158.0,-7.0){\makebox(0,0)[c]{$|x|$}}
\put(-408.0,67.0){\makebox(0,0)[c]{$v$}}
\put(-116.0,69.0){\makebox(0,0)[c]{$\tfrac{1}{2}$}}
\put(-13.0,-7.0){\makebox(0,0)[c]{$|x|$}}
\put(-303.0,-7.0){\makebox(0,0)[c]{$|x|$}}
\put(-323.0,57.0){\makebox(0,0)[c]{$|\Phi|$}}
\put(-181.0,57.0){\makebox(0,0)[c]{$g$}}
\put(-22.0,54.0){\makebox(0,0)[c]{$h$}}
\end{picture}
 \caption{The qualitative radial behavior of the scalar field $\Phi$ and the functions $g$ and $h$ appearing in the potentials $B$ and $A$ for a self-dual string solution \eqref{eq:SDS} with any of the $f(\xi)$ in \eqref{eq:choices_f}. Note that $g(\xi)\sim \sqrt{\xi}$ for large $\xi$.}
 \label{fig:SDSA4solution}
\end{figure}

The above ansatz reduces the self-dual string equation to the following ODEs:
\begin{equation}
\begin{aligned}
f(\xi)+\tfrac12 g(\xi)-g(\xi)h(\xi)-\xi f'(\xi)&=0~,\\
f(\xi)-2f(\xi)h(\xi)-\tfrac23\xi g'(\xi)&=0~.
\end{aligned}
\end{equation}
Note that $h(\xi)$ appears only algebraically. Assuming that $g(\xi)$ vanishes only at isolated points, we can combine the above equations into a single ODE for $f(\xi)$ and $g(\xi)$:
\begin{equation}
 f(\xi)^2-\xi f(\xi)f'(\xi)+\tfrac{1}{3}\xi g(\xi)g'(\xi)=0~.
\end{equation}
The fact that we arrive at a single ODE for two functions shows that our ansatz was underconstraint. This gives us the freedom to choose a function $f$ such that $\Phi$ has the correct asymptotic behavior $|\Phi|\sim v-|x|^{-2}$, which implies $f(\xi)\sim \xi$ at infinity. Convenient choices satisfying this property are e.g.\
\begin{equation}\label{eq:choices_f}
\begin{aligned}
f(\xi)&=\xi\ {\rm coth}(\xi)-1~,\\
f(\xi)&=\xi-1+\frac{2}{\pi  }\tan^{-1}\left(\frac{2}{\pi \xi}\right)~,\\
f(\xi)&= \xi\left(1-\frac{1}{1+\xi}\right)~.
\end{aligned}
\end{equation}
Moreover, we can choose an initial value for $g$ such that $g(0)=0$. The analytical expressions for $g(\xi)$ and $h(\xi)$ can be computed, but their analytical form does not provide further insight. For example, for the third choice in \eqref{eq:choices_f}, we have
\begin{equation}
 \begin{aligned}
  g(\xi)&=\sqrt{15+6\left(\xi-\frac{5+6\xi}{2(1+\xi)^2}-3 \log(1+\xi)\right)}~,\\
  h(\xi)&=\frac{1}{2}-\frac{\xi^2}{3(1+\xi)\sqrt{\tfrac{1}{3}\xi(6+\xi(9+2\xi))-2(1+\xi)^2\log(1+\xi)}}~.
 \end{aligned}
\end{equation}
The qualitative behavior resulting from any of the choices for $f(\xi)$ is displayed in figure \ref{fig:SDSA4solution}.
 
\subsection{Matrix representation of $A_4$ and hermitian 3-algebras}

The differential crossed module $A_4\rightarrow \frg_{A_4}$ can be represented in terms of matrices in the following way:
\begin{equation}
e_\mu:=\frac{1}{\sqrt{2}}\left(\begin{array}{cc} 0 & \sigma_\mu\\ 0 & 0\end{array}\right)\eand
D(e_\mu,e_\nu):=\frac{1}{2}\gamma_5\gamma_{\mu\nu}=\frac{1}{2}\left(\begin{array}{cc}\sigma_{\mu\nu} & 0\\ 0 & -\sigmab_{\mu\nu} \end{array}\right)~,
\end{equation}
where $\sigma_{\mu\nu}:=\sigma_{[\mu}\sigma^\dagger_{\nu]}$ and $\sigmab_{\mu\nu}:=\sigma^\dagger_{[\mu}\sigma_{\nu]}$. Note that we always use weighted antisymmetrization of indices.  The commutator in $\frg_{A_4}$ and the action $\acton$ of $\frg_{A_4}$ onto $A_4$ are just the matrix commutator.

In this notation, the solution \eqref{eq:SDS} becomes
\begin{equation}\label{eq:SDSsol2}
\begin{aligned}
\Phi&=\frac{1}{\sqrt{2}}\left(\begin{array}{cc} 0 & \frac{x}{|x|^3}\\ 0 & 0\end{array}\right)f(\xi)~,\\
B&=-\frac{1}{\sqrt{2}^3|x|^3}\left(\begin{array}{cc} 0 & x\,\dd \xb\wedge\dd x+\dd x\wedge \dd \xb\, x\\ 0 & 0\end{array}\right)g(\xi)~,\\
A&= \frac{1}{2|x|^2}\Im\left(\begin{array}{cc}x\,\dd \xb & 0\\ 0 & -\xb\,\dd x \end{array}\right)h(\xi)~,
\end{aligned}
\end{equation}
and the self-dual string equation becomes
\begin{equation}
H:=\dd B+[A, B]=\star(\dd\Phi+[A, \Phi])~.
\end{equation}

Interestingly, we see that the gauge potential, up to its radial behavior, is a combination of an instanton and an anti-instanton for gauge group $\sSU(2)$. 

Just as an $\sSU(2)$ monopole can be embedded into gauge groups with larger rank to obtain more general instanton solutions, we can embed our $A_4$ self-dual string solution into matrix representations of more general hermitian 3-algebras. These are given by differential crossed modules of the form
\begin{equation}
\begin{aligned}
\left(\begin{array}{cc} 0 &\cdot\\ 0 & 0\end{array}\right)\overset{\sft}{\rightarrow}\left(\begin{array}{cc} \cdot &0\\ 0 & \cdot\end{array}\right)
\end{aligned}
\end{equation}
with trivial map $\sft$, see appendix \ref{app:diff} for definitions. In particular, the 3-algebras appearing in the ABJM-model for $N$ M2-branes can be viewed in this way by considering blocks of $N\times N$-dimensional matrices. The off-diagonal block is given by elements of $\asu(N)\oplus \di \au(1)$, while the blocks on the diagonal form elements of $\au(N)$, cf.\ \cite{Palmer:2013ena}.

\subsection{Topological charges}\label{sec:top}

Similarly to the case of magnetic monopoles we may set the asymptotic value of the scalar field $\Phi$ for an $A_4$ self-dual string to a specific matrix, up to a gauge transformation\footnote{A detailed discussion of gauge transformation in higher gauge theory is postponed to section \ref{sec:fakecurvature}.}
\begin{equation}\label{eq:asymptotic}
\begin{aligned}
\Phi\sim \gamma\acton\frac{1}{\sqrt{2}}\left(\begin{array}{cccc} 0 &0&\di v&0\\0&0& 0 & \di v\\0&0&0&0\\0&0&0&0\end{array}\right)~~\mbox{as}~|x|\rightarrow\infty~,
\end{aligned}
\end{equation}
where $\gamma\in\sSU(2)\times\sSU(2)$ defines a map: $S^3_\infty\rightarrow \sSU(2)\times\sSU(2)/\sSU(2)$. The stabilizing group $\sSU(2)$ in the denominator is the unbroken symmetry group which leaves the form of $\Phi$ invariant. Since 
\begin{equation}
\begin{aligned}
\pi_3\left( \sSU(2)\times\sSU(2)/\sSU(2)\right)\cong\RZ~,
\end{aligned}
\end{equation}
we find that $A_4$ self-dual strings are indeed classified by an integer charge.

The element $\gamma\in\sSU(2)\times\sSU(2)/\sSU(2)$ in \eqref{eq:asymptotic} that produces solution \eqref{eq:SDSsol2} can be represented by
\begin{equation}
 \gamma=\left(\begin{array}{cc} x & 0 \\ 0 & 1 \end{array}\right)~,
\end{equation}
and we see that our solution has indeed charge 1, as the map $x:\sSU(2)\rightarrow \sSU(2)$ has winding number 1.

A charge formula analogue to \eqref{eq:top_charge_monopole} reads as
\begin{equation}
 (2\pi)^3 q=\tfrac{1}{2}\int_{S^3_\infty}\frac{(H,\Phi)}{||\Phi||}\ewith ||\Phi||:=\sqrt{\tfrac{1}{2}(\Phi,\Phi)}~,
\end{equation}
where $(\cdot,\cdot)$ denotes the Euclidean inner product on $A_4\cong \FR^4$. As the solutions arising from any of our choices \eqref{eq:choices_f} have all the same asymptotic behavior, they all yield the same result $q=1$.

\subsection{Comments on the solution}\label{ssec:comments}

To view this solution as a solution to the $\CN=(1,0)$ BPS equation \eqref{eq:(1,0)BPS}, we embed the gauge field $A$ taking values in $\frg_{A_4}$ into $A_4\oplus\frg_{A_4}$ and set the map $\sfd:(A_4\oplus\frg_{A_4})\odot(A_4\oplus\frg_{A_4})\rightarrow A_4$ to 
\begin{equation}
\sfd\left(\binom{a_1}{b_1},\binom{a_2}{b_2}\right)=\tfrac{1}{2}(b_1\acton a_2+b_2\acton a_1)~,
\end{equation}
for $a_{1,2}\in A_4,~b_{1,2}\in\frg_{A_4}$,  cf.\ \cite{Palmer:2013pka}. Equation \eqref{eq:(1,0)BPS} then reduces to the self-dual string equation \eqref{eq:NonAbSelfDualString}.

We may now look at the equation of motion
\begin{equation}
\nabla^2\Phi=\star\sfd(\CF,\star\CF)~.
\end{equation}
Our ansatz alone implies $F\acton\star B=0$ and so the equation of motion reduces to $\nabla^2\Phi=0$, which also appears in the Lambert-Papageorgakis equation of motion \eqref{eq:(2,0)eom}. Unfortunately, imposing $\nabla^2\Phi=0$ yields
\begin{equation}
 f(\xi)=\xi~,~~~g(\xi)=0~,~~~h(\xi)=\tfrac{1}{2}~.
\end{equation}
This solution does not have the desired behavior at $\xi=0$ and $\xi=\infty$. Moreover, the field strength $H=\star \nabla \Phi$ vanishes.

To solve this issue, note that from the higher gauge theory point of view, the condition $\nabla^2\Phi=0$ should in fact be ``categorified'' to 
\begin{equation}
 \nabla^2\Phi=\sft(\star\{B,B\})~,
\end{equation}
where the right hand side will become clear from the discussion in section \ref{sec:fakecurvature}. This condition is automatically satisfied by our ansatz.

We also note that the gauge transformations of \cite{Samtleben:2011fj} allow for components of $A$ to be turned on in $A_4$, which in turn introduces the terms in $H$ involving only $A$ in \eqref{eq:(1,0)BPS}. Exactly the same structure appears in the modified non-abelian gerbes of \cite{Ho:2012nt}, where the fake curvature condition was also dropped.

\section{Higher instantons}

\subsection{Preliminaries}

We define a \emph{higher instanton} as a solution to the six-dimensional self-duality equation
\begin{equation}\label{eq:self-duality_6d}
H=\star H~,~~~H:\dd B+A \acton B~,
\end{equation}
on $\FR^{1,5}$, where $A$ and $B$ are potential one- and two-forms taking values in the Lie algebras of a differential crossed module as before. We furthermore require that the curvature $H$ vanishes as $|x|\rightarrow\CCI$, implying that the solution extends to the conformal compactification of $\FR^{1,5}$. Here, $\CCI$ denotes the boundary of Minkowski space also known as {\em conformal infinity}, consisting of space-like, time-like and light-like infinity, see e.g.\ \cite{Penrose:1986ca} for more details. Comparing with the BPST instanton, we therefore expect that $H$ comes with a coefficient $\frac{1}{(\rho^2+|x|^2)^n}$ with $n\geq 2$.

Because we are dealing with a space with indefinite signature, we cannot expect our solutions to be regular everywhere. The fall-off behavior requires to include the norm of $x\in \FR^{1,5}$, and the expected coefficient $\frac{1}{(\rho^2+|x|^2)^n}$ therefore will yield divergences on a hyperboloid in $\FR^{1,5}$. In a neighborhood of the origin, however, the solutions will remain non-singular. In principle, we could apply a Wick rotation to $\FR^6$, but this would yield complex solutions of $H=\star\di H$.

Solutions to equations closely related to \eqref{eq:self-duality_6d} were previously constructed in \cite{Chu:2013joa}. These equations were interpreted as M-waves and the curvature of the solution's gauge potential one-form was given by an instanton solution.

We will now follow our strategy for self-dual strings and try to find as close an analogue to the BPST solution of instantons as possible. 

\subsection{Elementary higher instanton}

In section \ref{sec:MI}, we saw how the expression
\begin{equation}
\frac{\dd x\wedge\dd \xb}{(1+|x|^2)^2}
\end{equation}
appears both in the radially independent part of the Dirac monopole, where $x$ is the complex coordinate on one patch of $\CPP^1$, as well as in the basic instanton, where $x$ is a quaternionic coordinate on one patch of $S^4\cong \FH P^1$. This expression also describes a so-called octonionic instanton on $\FR^8$ when $x$ is an octonion \cite{Fubini:1985jm}. In this section, we use the analogous self-dual three-forms on $\FR^{1,5}$ to find solutions to the higher instanton equations.

We denote the van-der-Waerden symbols appearing in the Clifford algebra of $\FR^{1,5}$ by $\sigma_M$, $M=0,\ldots,5$. We use the representation given implicitly by
\begin{equation}
  x_{AB}\ =\ x_M\sigma^M_{AB}\ =\ \left(\begin{array}{cccc}
			  0 & x_0+x_5 & -x_3-\di x_4 & -x_1+\di x_2\\
			  -x_0- x_5 & 0 & -x_1-\di x_2 & x_3-\di x_4\\
			  x_3+\di x_4 & x_1+\di x_2 & 0 & -x_0+x_5\\
			  x_1-\di x_2 & -x_3+\di x_4 & x_0-x_5 & 0
			\end{array}\right)~.
\end{equation}
We also define
\begin{equation}
 \xh=(\xh^{AB}):=(\tfrac{1}{2}\eps^{ABCD}x_{CD})~.
\end{equation}
We then have $x^\dagger=-\xh$ and $\xh^\dagger=-x$ and the norm of the vector $x$ is given by
\begin{equation}
 |x|^2=-\tfrac{1}{4}\tr(\xh x)=\sqrt{\det(x)}=\sqrt{\det(\xh)}~.
\end{equation}
Note also that
\begin{equation}
 x^{-1}=\frac{-\xh}{|x|^2}\eand \xh^{-1}=\frac{-x}{|x|^2}~.
\end{equation}
With this convention, the three-forms $\dd\xh\wedge\dd x\wedge\dd \xh$ and $\dd x\wedge\dd\xh\wedge\dd x$ are self-dual and anti-self-dual, respectively. 

As differential crossed module, we consider $\frh\stackrel{\sft}{\rightarrow}\frg$ with $\frh\cong \FR^{1,15}\supset \FR^{1,5}$ and $\frg=\aspin(1,5)$. We use a matrix representation similar to that for $A_4$. That is, we work with block matrices
\begin{equation}
 \left(\begin{array}{cc}M_1 & M_2\\ 0 & M_3\end{array}\right)~,
\end{equation}
where the $M_i$ are $4\times 4$-dimensional complex matrices. Elements of $\frg$ have $M_2=0$ and elements of $\frh$ have $M_1=M_3=0$. 

A first abelian solution of the self-duality equation \eqref{eq:self-duality_6d}, which is singular at the origin $x=0$, is given by the following fields:
\begin{equation}\label{eq:sing}
A=0~,~~~B=\tfrac{\rho^3}{|x|^6}x~\dd  \xh\wedge\dd x~,~~~H=\dd B = \tfrac{\rho^3}{|x|^8}x~ \dd \xh\wedge\dd x\wedge\dd \xh~x~.
\end{equation}

To find true non-abelian solutions of the form $H\sim \dd\xh\wedge\dd x\wedge\dd \xh$ with the right fall-off behavior, we make the following ansatz for the $B$-field:
\begin{subequations}
\begin{equation}
 B=\frac{1}{(\rho^2+|x|^2)^\frac{3}{2}}\left(\begin{array}{cc} 0 &\xh\,\dd x\wedge\dd \xh-\dd\xh\, x\wedge\dd \xh+\dd\xh\wedge \dd x\, \xh\\ 0 & 0\end{array}\right)~.
\end{equation}
Here, the power of the fall-off coefficient $\frac{1}{(\rho^2+|x|^2)}$ is determined by the fact that $B$ has to be dimensionless. Together with the instanton-inspired gauge potential
\begin{equation}
 A=\frac{1}{4(\rho^2+|x|^2)}\left(\begin{array}{cc} \dd \xh\,x-\xh\, \dd x & 0\\ 0 & \dd x\,\xh-x\, \dd \xh\end{array}\right)~,
\end{equation}
we obtain the self-dual three-form curvature
\begin{equation}
 H:= \frac{\rho^2}{(\rho^2+|x|^2)^\frac{5}{2}}\left(\begin{array}{cc} 0 &\dd \xh\wedge\dd x\wedge\dd \xh\\ 0 & 0\end{array}\right)=\star H
\end{equation}
as well as the two-form curvature
\begin{equation}
 F=-\frac{1}{(\rho^2+|x|^2)^2}\left(\begin{array}{cc} \rho^2\,\dd \xh\wedge \dd x+\tfrac{1}{2} \dd \xh\, x\wedge \dd \xh\, x & 0\\ 0 & \rho^2\,\dd x\wedge\dd\xh+\tfrac{1}{2}\dd x\,\xh\wedge \dd x\, \xh\end{array}\right)~.
\end{equation}
\end{subequations}

\subsection{Comments on the higher instanton solution}

As the coefficients controlling the fall-off appear with non-integer powers in $B$ and in particular in $H$, the above solution is only defined for $\rho^2+|x|^2>0$, i.e.\ in the region of $\FR^{1,5}$ containing the origin, which is bounded by the hyperboloid $\rho^2+|x|^2=0$. On the hyperboloid itself, the solution blows up, as expected. Outside of the hyperboloid, the above solution is purely imaginary. Multiplying it by an appropriate root of $-1$ then turns it again into a real solution. 

Note that because of the fall-off behavior of our solution, it extends to the region of the conformal compactification of Minkowski space that consists of the interior of the hyperboloid $\rho^2+|x|^2=0$.

Imposing less stringent conditions on the shapes of $A$, $B$, $H$ and $F$, many more general solutions can be found. In particular, one can replace the antisymmetrizations in the potential one- and two-forms, such as $\dd \xh\,x-\xh\,\dd x$, by more general terms, such as $\alpha_1\dd \xh\,x-\alpha_2\xh\,\dd x$ with constants $\alpha_{1,2}\in \FC$. Self-duality of $H$ then does not fix all the arising constants. The resulting curvatures $H$ and $F$, however, look less natural or symmetric.

Moreover, one easily realizes that our solutions can be `conjugated' to anti-higher instanton solutions satisfying $H=-\star H$. Explicitly, one needs to take the conjugate transpose and apply time-reversal on the fields.

\section{Fake curvature, gauge transformations and differential 2-crossed modules}\label{sec:fakecurvature}

In this section, we address remaining questions related to gauge symmetry and the so-called fake curvature condition in higher gauge theory. Recall that the local description of a principal 3-bundle is given in terms of one-, two- and three forms $A$, $B$ and $C$ which take values in a differential 2-crossed module consisting of Lie algebras $\frg$, $\frh$ and $\frl$. The corresponding curvatures read as
\begin{equation}
  F:=\dd A+\tfrac{1}{2}[A,A]~,~~~H:=\dd B+A\acton B~,~~~G:=\dd C+A\acton C+ \{B,B\}~.
\end{equation}
Here, $\acton$ denotes the actions of $\frg$ onto $\frh$ and $\frl$, while $\{-,-\}$ is the so-called Peiffer lifting, cf.\ appendix \ref{app:diff}. It is well-known that for the parallel transport of a one-dimensional object along a surface to be invariant under reparameterizations of the surface, the so-called {\em fake curvature 2-form} has to vanish:
\begin{equation}\label{eq:fake_2_curvature}
 \CF:=F-\sft(B)=0~,
\end{equation}
cf.\ \cite{Baez:2004in}. Moreover, a consistent parallel transport of two-dimensional objects along a volume requires the {\em fake curvature 3-form} to vanish:
\begin{equation}\label{eq:fake_3_curvature}
 \CH:=H-\sft(C)=0~,
\end{equation}
cf.\ \cite{Saemann:2013pca} for a full and global description of the underlying Deligne cohomology.

The (finite) gauge transformations of the gauge potentials are given by \cite{Saemann:2013pca}
\begin{equation}\label{eq:gauge_transformations}
 \begin{aligned}
\tilde C&=\gamma^{-1}\acton C-(\tilde\nabla+\sft(\Lambda)\acton)\Sigma+\{\tilde B+\tfrac12\tilde{\nabla}\Lambda+\tfrac12[\Lambda,\Lambda],\Lambda\}+\{\Lambda,\tilde B-\tfrac12\tilde{\nabla}\Lambda-\tfrac12[\Lambda,\Lambda]\}~,\\
 \tilde B&=\gamma^{-1}\acton B-\tilde{\nabla}\Lambda-\tfrac12\sft(\Lambda)\acton \Lambda-\sft(\Sigma)~,\\
 \tilde A&=\gamma^{-1}A \gamma+\gamma^{-1}\dd \gamma-\sft(\Lambda)~,
 \end{aligned}
\end{equation}
where $\gamma$ is a function with values in a Lie group $\mathsf{G}$ with $\frg=\mathsf{Lie}(\mathsf{G})$, and $\Lambda$ and $\Sigma$ are a $\frh$-valued one-form and an $\frl$-valued two-form, respectively. We also used the abbreviation $\tilde \nabla\ :=\ \dd+\tilde A\acton$. For future reference, let us also note that the three-form curvature $H$ transforms according to
\begin{equation}\label{eq:H-trafo}
 \begin{aligned}
  \tilde H&= \gamma^{-1}\acton H-\big(\tilde F-\sft(\tilde B)\big)\acton\Lambda+\\
 &\hspace{0.4cm}+\, \sft\Big[-(\tilde\nabla+\sft(\Lambda)\acton)\Sigma+\{\tilde B+\tfrac12\tilde{\nabla}\Lambda+\tfrac12[\Lambda,\Lambda],\Lambda\}+\{\Lambda,\tilde B-\tfrac12\tilde{\nabla}\Lambda-\tfrac12[\Lambda,\Lambda]\}\Big]~.
 \end{aligned}
\end{equation}
The fake curvature condition \eqref{eq:fake_2_curvature} is covariant under gauge transformations and so is \eqref{eq:fake_3_curvature}, provided the fake curvature 2-form vanishes.

In the following, let us consider how our solutions fit into this framework. In the case of the self-dual string solution, we only ever parallel transport point-like objects. In particular, our self-dual string extends in the $x^5$ directions, cf.\ \eqref{diag:M2M5} and only its point-like position in $\FR^4$ is relevant in the self-dual string equation \eqref{eq:NonAbSelfDualString}. Therefore, it is not necessary to impose the fake curvature condition \eqref{eq:fake_2_curvature}.

There are, however, several advantages to implementing condition \eqref{eq:fake_2_curvature} anyway. First of all, this would restrict $A$ further and guarantee that the relevant degrees of freedom are all contained in $B$. Second, it would mean that the resulting solution is one of those arising in the twistor construction of \cite{Saemann:2012uq}. Finally, the fake curvature condition \eqref{eq:fake_2_curvature} leads to a Bianchi identity for $H$, which is very useful.

It is clear that to impose \eqref{eq:fake_2_curvature}, we have to generalize our differential crossed module $A_4\stackrel{\sft}{\rightarrow}\frg_{A_4}$, as the triviality of $\sft$ would imply $F=0$. Note that a higher gauge theory on principal $2$-bundles with non-vanishing fake curvature can be reformulated as a higher gauge theory on principal $3$-bundles with vanishing fake curvature \cite{Schreiber:N01,Baez:2010ya}. We used this idea previously to demonstrate that the ABJM model is a higher gauge theory \cite{Palmer:2013ena}.

Here, we can follow the same path as in \cite{Palmer:2013ena}. We replace our differential crossed module with the corresponding differential 2-crossed module of inner derivations $\frder(A_4\stackrel{\sft}{\rightarrow}\frg_{A_4})=A_4\stackrel{\sft}{\rightarrow}\frg_{A_4}\ltimes A_4\stackrel{\sft}{\rightarrow}\frg_{A_4}$, where in matrix representation the two maps $\sft$ read as 
\begin{equation}
\sft:\left(\begin{array}{cc} 0 & h\\ 0 & 0\end{array}\right)\mapsto \left(\begin{array}{cc} 0 & -h\\ 0 & 0\end{array}\right) \eand 
\sft:\left(\begin{array}{cc} g_L & h\\ 0 & g_R\end{array}\right)\mapsto \left(\begin{array}{cc} g_L & 0\\ 0 & g_R\end{array}\right)~,
\end{equation}
respectively, for $(g_L,g_R)\in \frg_{A_4}$ and $h\in A_4$. More details on this point are given in appendix \ref{app:diff}.

Starting from a solution $(A_0, B_0)$ to the self-dual string equation based on the differential crossed module $A_4\stackrel{\sft}{\rightarrow}\frg_{A_4}$, we obtain a solution based on the differential 2-crossed module $\frder(A_4\stackrel{\sft}{\rightarrow}\frg_{A_4})$ by letting $A=A_0$, $B=B_0+\dd A+\tfrac{1}{2}[A,A]$ and $C=\dd B_0+[A,B_0]$. Then both \eqref{eq:fake_2_curvature} and \eqref{eq:fake_3_curvature} are automatically satisfied.

Note, however, that the self-dual string equation is not invariant under the general gauge transformations \eqref{eq:gauge_transformations}. According to the results of \cite{Saemann:2012uq,Saemann:2013pca}, analogues of adjoint scalar fields in higher gauge theory such as $\Phi$ transform in the same way as the three-form curvature $H$. Moreover, for a covariant derivative to make sense, the possible gauge transformations have to restrict to $H\rightarrow \tilde{H}:=\gamma^{-1}\acton H$. This breaks the gauge symmetries \eqref{eq:gauge_transformations} to a residual symmetry given by triples $(\gamma,\Lambda,\Sigma)$ with
\begin{equation}\label{eq:residual_gauge}
 -(\tilde\nabla+\sft(\Lambda)\acton)\Sigma+\{\tilde B+\tfrac12\tilde{\nabla}\Lambda+\tfrac12[\Lambda,\Lambda],\Lambda\}+\{\Lambda,\tilde B-\tfrac12\tilde{\nabla}\Lambda-\tfrac12[\Lambda,\Lambda]\}=0~.
\end{equation}

This observation is crucial: Because of the simple structure of our differential crossed and 2-crossed module, the solution would be gauge trivial if the gauge symmetries \eqref{eq:gauge_transformations} were not broken. The fact that equations of motion break the general gauge symmetries of higher gauge theory\footnote{which one might regard as the larger gauge symmetries of a flat connective structure} seems not unusual and has been observed previously in \cite{Baez:2012bn} and \cite{Palmer:2013ena}. 

We thus arrive at a solution of the self-dual string equation based on a differential 2-crossed module satisfying both fake curvature conditions \eqref{eq:fake_2_curvature} and \eqref{eq:fake_3_curvature}. Such a solution should now have a twistor description in terms of holomorphic 3-bundles as described in \cite{Saemann:2013pca}.

In the case of the higher instantons, which should arise in a theory capturing the parallel transport of self-dual strings, we definitely do need to impose the fake curvature condition \eqref{eq:fake_2_curvature}. We can do this by applying precisely the same strategy as before. In particular, we can again replace the differential crossed module there with the corresponding differential 2-crossed module of inner derivations. To break the gauge symmetry here, we demand that $C=0$, as this field is not expected to be relevant, anyway. This again yields a residual gauge transformation parameterized by triples $(\gamma,\Lambda,\Sigma)$ satisfying \eqref{eq:residual_gauge}. We thus obtain a higher instanton solution where only the fake curvature condition \eqref{eq:fake_2_curvature} is imposed. This is sufficient, because the six-dimensional theory should describe the parallel transport of self-dual strings.

\section{Conclusions and future directions}

In this paper, we found solutions to the non-abelian self-dual string and higher instanton equations. These solutions were obtained from generalizing the 't Hooft-Polyakov monopole solution as well as the BPST instanton solution to higher gauge theory. 

Our self-dual string solution behaves in complete analogy to the 't Hooft-Polyakov mono\-pole. Its Higgs field is non-singular and exhibits the right fall-off behavior. It has a clearly identifiable unit topological charge. 

The gauge structure underlying the solution was given by a differential crossed module corresponding to $A_4$, which featured prominently in a recently popular M2-brane model \cite{Bagger:2007jr,Gustavsson:2007vu}. In particular, the same gauge algebra underlies the elementary solutions to the Basu-Harvey equation. This supports our conjecture \cite{Palmer:2012ya,Palmer:2013ena} that the description of the effective BPS regimes of M2- and M5-branes are very closely related. 

While the ansatz for the 't Hooft-Polyakov monopole gives a unique solution, this was not the case for our ansatz. This issue could be fixed by correlating the radial decay functions for the scalar field, the one- and the two-form potential. Alternatively, one might want to impose an additional condition on the two-form curvature. The twistor picture of self-dual string solutions obtained in \cite{Saemann:2012uq,Saemann:2013pca}, however, suggests that this is not the natural thing to do.

We also found solutions to the self-duality equation in six dimensions, generalizing the BPST instanton to higher gauge theory. This equation is necessarily discussed on a space with indefinite signature. Together with an expected fall-off behavior of our solutions towards conformal infinity, this requires them to blow up on the hyperboloid $\rho^2+|x|^2=0$. In the interior of the hyperboloid, however, our solutions were non-singular.

While our solutions do not satisfy the standard fake curvature condition of higher gauge theory on principal 2-bundles, we showed that they can be readily extended to corresponding connective structures on principal 3-bundles, which do satisfy the corresponding fake curvature conditions.

Our construction of these elementary solutions is a first step towards translating the Nahm transform to higher gauge theory, which is one of our key future goals. For this, it might also be necessary to construct further elementary solutions, which do satisfy all relevant fake curvature conditions without breaking any gauge symmetry. This should also constrain them further, yielding unique solutions. Such solutions have to be based on more complicated crossed modules\footnote{or even semi-strict Lie 2-algebras} than the rather trivial ones discussed in this paper. Most likely, one should study infinite dimensional crossed modules such as the ones constructed in \cite{Baez:2005sn}.

\vspace*{-0.3cm}

\section*{Acknowledgements}

We would like to thank David Berman, Neil Lambert and in particular Martin Wolf for discussions. This work was supported by the EPSRC Career Acceleration Fellowship EP/H00243X/1.

\appendices

\subsection{3-algebras}\label{app:3algebras}

In this paper, we use the term \emph{3-algebra} to refer collectively to \emph{real 3-algebras} and \emph{hermitian 3-algebras}. These are not to be confused with Lie 3-algebras.

A {\em 3-Lie algebra} \cite{Filippov:1985aa} is a real vector space $\CA$ endowed with a totally antisymmetric, trilinear 3-bracket $[-,-,-]:\CA^{\wedge 3} \rightarrow \CA$ satisfying the {\em fundamental identity}
\begin{equation}\label{eq:fundamentalIdentity}
 [a,b,[c,d,e]]=[[a,b,c],d,e]+[c,[a,b,d],e]+[c,d,[a,b,e]]
\end{equation}
for all $a,b,c,d,e\in\CA$. If we endow $\CA$ with a metric which satisfies the {\em compatibility condition}
\begin{equation}
 ([a,b,c],d)+(c,[a,b,d])=0~,
\end{equation}
we arrive at a {\em metric 3-Lie algebra}. An explicit example of a metric 3-Lie algebra is discussed at the beginning of section \ref{ssec:A4solSDS}.

A 3-Lie algebra $\CA$ always comes with an associated Lie algebra $\frg_\CA$ of inner derivations. The vector space of inner derivations is the linear span of $D(a,b)$, where 
\begin{equation}
 D(a,b)\acton c:=[a,b,c]~,~~~a,b,c\in\CA~.
\end{equation}
This forms a Lie algebra due to the fundamental identity \eqref{eq:fundamentalIdentity}.

A {\em real 3-algebra} \cite{Cherkis:2008qr} is a generalized 3-Lie algebra in which the ternary bracket is antisymmetric only in its first two slots.

On the other hand, a {\em hermitian 3-algebra} \cite{Bagger:2008se} is a complex vector space $\CA$ endowed with a 3-bracket $[-,-;-]$ which is linear and antisymmetric in its first two slots and antilinear in its third slot and satisfies the {\em fundamental identity} 
\begin{equation}\label{eq:h3fundamentalIdentity}
 [[a,b;c],d;e]=[[a,d;e],b;c]+[a,[b,d;e];c]-[a,b;[c,e;d]]
\end{equation}
for all $a,b,c,d,e\in\CA$. Together with a hermitian form satisfying the {\em compatibility condition}
\begin{equation}
([a,b;c],d)=(b,[c,d;a])~,
\end{equation}
we have a {\em metric hermitian 3-algebra}.

Analogously to a real 3-algebra, a hermitian 3-algebra $\CA$ comes with a Lie algebra of inner derivations $\frg_\CA$, spanned by $D(a,b)$ with $D(a,b)\acton c := [c,a;b]$ for $a,b,c\in \CA$. 

\subsection{Differential 1- and 2-crossed modules}\label{app:diff}

As shown in \cite{Palmer:2012ya}, metric real and hermitian 3-algebras form special examples of differential crossed modules, which play an important role in higher gauge theory.

A \emph{differential crossed module} consists of a pair of Lie algebras $\frg$, $\frh$ together with an action $\acton$ of $\frg$ onto $\frh$ as derivations and a Lie algebra homomorphism: $\sft:\frh\rightarrow\frg$. The maps $\acton$ and $\sft$ satisfy the following relations:
\begin{equation}\label{eq:Pfeif}
 \sft(g\acton h)=[g,\sft(h)]\eand \sft(h_1)\acton h_2=[h_1,h_2]
\end{equation}
for all $g\in \frg$ and $h\in \frh$. The first identity is an equivariance condition while the second is known as the \emph{Peiffer identity}. 

This structure, when endowed with metrics on $\frg$ and $\frh$, contains the well known metric 3-algebras relevant to M2-brane models via the Faulkner construction \cite{Palmer:2012ya}. Explicitly, the metrics on $\frg$ and $\frh$ lead to a triple bracket  
\begin{equation}\label{eq:metric}
 ( D(a_1,a_2),D(a_3,a_4))_\frg=(D(a_1,a_2)\acton a_3,a_4)_\frh=([a_1,a_2,a_3],a_4)_\frh~.
\end{equation}
Inversely, a triple bracket can be used to define a unique premetric on $\frg$.

A \emph{differential 2-crossed module} is defined as a complex of Lie algebras
\begin{equation}
 \frl\ \xrightarrow{~\sft~}\ \frh\ \xrightarrow{~\sft~}\ \frg
\end{equation}
along with $\frg$-actions $\acton$ on $\frh$ and $\frl$ by derivations and a $\frg$-equivariant bilinear map, called {\em Peiffer lifting}: $\{\cdot,\cdot\}: \frh\times \frh\rightarrow \frl$, which encodes the failure of the Peiffer identity \eqref{eq:Pfeif} to hold. These maps satisfy the following axioms:
\begin{conditions}
 \item[(i)] $\sft(g\acton l)=g\acton\sft(l)$ and $\sft(g\acton h)=[g,\sft(h)]$ ,
 \item[(ii)] $\sft(\{h_1,h_2\})=[h_1,h_2]-\sft(h_1)\acton h_2$,
 \item[(iii)] $\{\sft(l_1),\sft(l_2)\}=[l_1,l_2]$,
 \item[(iv)] $\{[h_1,h_2],h_3\}=\sft(h_1)\acton\{h_2,h_3\}+\{h_1,[h_2,h_3]\}-\sft(h_2)\acton\{h_1,h_3\}-\{h_2,[h_1,h_3]\}$,
 \item[(v)] $\{h_1,[h_2,h_3]\}=\{\sft(\{h_1,h_2\}),h_3\}-\{\sft(\{h_1,h_3\}),h_2\}$ ,
 \item[(vi)] $\{\sft(l),h\}+\{h,\sft(l)\}=-\sft(h)\acton l$ ,
\end{conditions}
for all $g\in\frg$, $h\in\frh$, and $l\in\frl$, where $[\cdot,\cdot]$ denotes the Lie bracket in the respective Lie algebra. For more details, see e.g.\ \cite{Saemann:2013pca}.

An example of such a structure relevant to our discussion is the following: Consider a differential crossed module $\frh\xrightarrow{~\tilde{\sft}~}\frg$ with action $\tilde{\acton}:\frg\times \frh\rightarrow \frh$. The differential 2-crossed module of inner derivations of $\frh\xrightarrow{~\tilde{\sft}~}\frg$, denoted $ \frder\big(\frh\xrightarrow{~\tilde{\sft}~}\frg\big)$, has the following underlying normal complex, cf. \cite{Roberts:0708.1741,Palmer:2013ena}:
\begin{equation}
\frh\ \xrightarrow{~\sft~}\ \frg \ltimes \frh\ \xrightarrow{~\sft~}\ \frg~.
\end{equation}
The two maps $\sft$ and the two $\frg$-actions are defined as
\begin{equation}
\begin{aligned}
 \sft(h):=(\tilde{\sft}(h),-h)&\eand \sft(g,h):=\tilde{\sft}(h)+g~,\\
 g\acton h := g~\tilde{\acton}~h&\eand g_1 \acton (g_2,h):=([g_1,g_2],g_1~\tilde{\acton}~h)~,
 \end{aligned}
\end{equation}
which yields the Peiffer lifting
\begin{equation}
 \{(g_1,h_1),(g_2,h_2)\}:=g_2\tilde{\acton} h_1
\end{equation} 
for $g,g_1,g_2\in\frg$, $h,h_1,h_2\in \frh$.

\bibliography{bigone2}
\bibliographystyle{latexeu}

\end{document}